\documentclass{pasa}%
\usepackage{anysize}
\usepackage[T1]{fontenc}
\usepackage{graphicx}

\usepackage{array}

\usepackage{aas_macros}
\usepackage{grffile}
\usepackage{amsmath}

\usepackage{bm}

\usepackage{url}
\usepackage{amsfonts}
\usepackage{lmodern}
\usepackage{algpseudocode}
\usepackage{algorithmicx}
\usepackage{algorithm}
\usepackage{xfrac}
\usepackage{float}
\usepackage{multirow}
\usepackage{balance}
\usepackage{tikz}

\algblock[Block]{Block}{EndBlock}
\algblockdefx[Block]{Block}{EndBlock}%
[1]{{#1}}%
[1]{{#1}}

\newcommand{\Given}[1]{\State{\bf given} {#1}}

\newcommand{\RepeatFor}[1]{\Repeat {\bf~for} {#1}}

\newcommand{\ignore}[1]{\color{magenta}\color{black}}

\title[Implementation of the $w$-stacking $w$-projection hybrid algorithm]{$w$-stacking $w$-projection hybrid algorithm for wide-field interferometric imaging: implementation details and improvements}

\author[Pratley et al.]{L. Pratley$^1$\thanks{Luke.Pratley@gmail.com}, 
M. Johnston-Hollitt$^2$ and J. D. McEwen$^1$
\affil{$^1$Mullard Space Science Laboratory (MSSL), University College London (UCL), Holmbury St Mary, Surrey RH5 6NT, UK}%
\affil{$^2$International Centre for Radio Astronomy Research (ICRAR)- Curtin University, 1 Turner Ave, Bentley, 6102, WA, Australia}
}%

\jid{PASA}
\doi{10.1017/pas.\the\year.xxx}
\jyear{\the\year}

\usepackage{aas_macros}
\usepackage{hyperref} 
\hypersetup{colorlinks,citecolor=blue,linkcolor=blue,urlcolor=blue}

\hypersetup{draft}
\begin{document}

\begin{frontmatter}
\maketitle

\begin{abstract}
We present a detailed discussion of the implementation strategies for a recently developed $w$-stacking $w$-projection hybrid algorithm used to reconstruct wide-field interferometric images. In particular, we discuss the methodology used to deploy the algorithm efficiently on a supercomputer via use of a Message Passing Interface (MPI) $k$-means clustering technique to achieve efficient construction and application of non co-planar effects. Additionally, we show that the use of conjugate symmetry increases the algorithms performance by imaging an interferometric observation of Fornax A from the Murchison Widefield Array (MWA). We perform exact non-coplanar wide-field correction for 126.6 million visibilities using 50 nodes of a computing cluster. The $w$-projection kernel construction takes only 15 minutes, demonstrating that the implementation is both fast and efficient.
\end{abstract}

\end{frontmatter}

\section{INTRODUCTION}
\label{sec:intro}
The advent of wide-field interferometers such as the Murchison Widefield Array \citep[MWA;][]{tin13}, Long Wavelength Array \citep[LWA;][]{ell09} and the Low Frequency Array \citep[LOFAR;][]{vH13} has created a number of imaging challenges. These challenges include the large number of measurements in each observation, the instrumental effects that are measurement dependent, and the large image sizes due to high resolution and wide-field of view. Additionally, these telescopes have a variety of science goals, including high priority science such as probing Galactic and extra-galactic magnetic fields (especially in low mass galaxy clusters; \citealt{mjh15}), and detecting the redshifted 21cm spectral line of the Epoch of Reionoization \citep{koo15}. Furthermore, the wide-field of view provides the advantage of observing many objects in a single pointing, reducing the observation time needed to survey the radio sky. If the imaging challenges are overcome, it will herald an era of unprecedented sensitivity and resolution for the low frequency sky, over extremely wide-field of views. 
%The imaging challenges of low frequency widefield interfermeters needs to be overcome for these science goals to be achieved. 

Non-coplanar baselines, $(u, v, w)$, in the presence of wide-fields of view produce measurement dependent effects, i.e.\ a directional dependent effect (DDE) that is different for each measurement. Each $w$ value provides a unique linear chirp that needs to be modelled in the image domain and applied during image reconstruction. Previously only groups of $w$ values have been corrected for \citep{Cornwell08,off14} when dealing with a standard observation. This has been through the use of two algorithms, the $w$-stacking algorithm, where average $w$ corrections are applied in the image domain to groups of measurements, and the $w$-projection algorithm, where average $w$-corrections are applied when degridding in the $(u, v, w)$ domain. The $w$-stacking algorithm has the trade off that a Fast Fourier Transform (FFT) needs to be applied for each $w$ group. The $w$-projection algorithm has the trade off that kernel construction can be expensive and the support size is large for large $w$ values. Both algorithms have been limited to correcting individual groups of measurements for large data sets.

Two recent developments have allowed individual correction for each data set. The first is the use of adaptive quadrature and radial symmetry to calculate $w$-projection kernels orders of magnitude faster than the full 2d calculation (\citealp{pra19b}, hereafter Paper I). The second is the developments in distributed image reconstruction from state of the art convex optimization algorithms, which provide a natural framework for the Message Passing Interface (MPI) distribution of FFTs and degridding for radio interferometric imaging \citep{pra19c}.  Recently, an MPI hybrid $w$-stacking $w$-projection algorithm demonstrating these developments was applied on a super computing cluster, where 17.5 million
 measurements were individually corrected over a 25 by 25 degree field of view from an MWA observation (Paper I). Such individual correction has not been previously possible.
 
After reviewing the $w$-stacking $w$-projection algorithm, we provide the algorithmic details of how to distribute the measurements through a $k$-means clustering algorithm to improve computational performance, the use of conjugate symmetry to reduce the range of $w$ values, and show the application of these algorithms to a larger data set to demonstrate the improvement. We end with a discussion of future strategies for kernel calculation and adapting the algorithm to model other DDEs.

The paper is laid out as follows. Section \ref{sec:radio_measurement_equation} introduces the wide-field interferometric measurement equation. Section \ref{sec:MPI_algo} describes the distributed $k$-means clustering algorithm used to create the $w$-stacks and the reconstruction algorithm used to generate a sky model of the observed data. Section \ref{sec:fornaxa} demonstrates the application of the algorithm for this implementation on an observation of Fornax A. Section \ref{sec:improvements} proposes possible improvements in kernel calculation for large data sets, and discusses how other directional dependent effects can be included into the algorithm. The work is concluded in Section \ref{sec:conclusion}.

\section{Wide-field Imaging Measurement Equation}					
\label{sec:radio_measurement_equation}
The non-coplanar wide-field interferometric measurement equation is
\begin{equation}
\begin{split}
	y(u, v, w^\prime) = \int x(l, m) a(l, m)\frac{{\rm e}^{-2\pi i w^\prime(\sqrt{1 -l^2 - m^2} - 1)}}{\sqrt{1 -l^2 - m^2}}\\
	\times {\rm e}^{-2\pi i (lu + mv)}\,  {\rm d}l{\rm d}m\, ,
	\end{split}
	\label{eq:meas_eq}
\end{equation}
where $(u, v, w^\prime)$ are the baseline coordinates and $(l, m, n)$ are directional cosines restricted to the unit sphere. In this work, we define $w^\prime = w + \bar{w}$, where $\bar{w}$ is the average value of $w$-terms, and $w$ is the effective $w$-component (with zero mean),
$x$ is the sky brightness and $a$ includes direction dependent effects such as the primary beam. The measurement equation is a mathematical model of the measurement process, i.e.\ signal acquisition, that allows one to calculate model measurements $y$ when provided with a sky model $x$. 

A number of methods can be used to solve for $x$ given samples $y$, such as CLEAN \citep{hog74}, Maximum Entropy \citep{Ables74,Cornwell85}, and Sparse Regularization algorithms \citep{mce10,ono16,LP18,da18,pra19b,pra19c}. Ultimately, all interferometric measurement equations are derived from the van Cittert-Zernike theorem \citep{zer38} and the measurement equation can be extended to include general direction dependent effects and polarization, and to solve for $x$ natively on the sphere \citep{mce08, smi11, pri15}. 
																						
To make use of the FFT, the measurement equation is traditionally calculated and approximated using degridding \citep{fes03,tho08}. The measurement equation can be represented by the following linear operations 
\begin{equation}
	\bm{y} = \bm{\mathsf{W}}\bm{\mathsf{G}}\bm{\mathsf{C}}\bm{\mathsf{F}}\bm{\mathsf{Z}}\bm{\mathsf{S}} \bm{x}\, .
	\label{eq:matrix_equation}
\end{equation}
$\bm{\mathsf{S}}$ represents a gridding correction and correction of baseline independent effects such as $\bar{w}$, $\bm{\mathsf{Z}}$ represents zero padding of the image, $\bm{\mathsf{F}}$ is an FFT, $\bm{\mathsf{G}}$ represents a sparse circular convolution matrix that interpolates measurements off the grid and the combined $\bm{\mathsf{G}}\bm{\mathsf{C}}$ includes baseline dependent effects such as variations in the primary beam and $w$-component in the interpolation, and $\bm{\mathsf{W}}$ are weights applied to the measurements. This linear operator is typically called a measurement operator $\bm{\mathsf{\Phi}} = \bm{\mathsf{W}}\bm{\mathsf{G}}\bm{\mathsf{C}}\bm{\mathsf{F}}\bm{\mathsf{Z}}\bm{\mathsf{S}}$ with $\bm{\mathsf{\Phi}} \in \mathbb{C}^{M \times N}$. Furthermore, $\bm{x}_i = x(\bm{l}_i)$ and $\bm{y}_k = y(\bm{u}_k)$ are discrete vectors in $\mathbb{R}^{N \times 1}$ and $\mathbb{C}^{M \times 1}$ in this setting.	The measurement operator has an adjoint operator $\bm{\mathsf{\Phi}}^\dagger$.  The dirty map can be calculated by $\bm{\mathsf{\Phi}}^\dagger\bm{y}$, and the residual map by $\bm{\mathsf{\Phi}}^\dagger\bm{\mathsf{\Phi}}\bm{x} - \bm{\mathsf{\Phi}}^\dagger\bm{y}$.

\section{Distributed Wide-Field Imaging}
\label{sec:MPI_algo}
In this section, we briefly describe the algorithmic details for the distributed $w$-projection $w$-stacking hybrid algorithm. 
																	
We use the interferometric image reconstruction software package PURIFY\footnote{\url{https://github.com/astro-informatics/purify}} (version 3.0.1, \citealt{purify}) developed in C++ \citep{car14,LP18,pra19c}, where the authors have implemented an MPI distributed measurement operator. The authors have also developed MPI distributed wavelet transforms, along with MPI variations of the alternating direction method of multipliers (ADMM) algorithm in the software package SOPT\footnote{\url{https://github.com/astro-informatics/sopt}} (version 3.0.1, \citealt{sopt}).

This is not the first time sparse image reconstruction has been used for wide-fields of view. In particular, the $w$-term is known to spread information across visibilities, increasing the effective bandwidth in what is known as the spread spectrum effect \citep{wia09,mce10,wol13,dab17}, increasing the possible resolution of the reconstructed sky model. But these previous works have been restricted to proof-of-concept studies. One of the advantages of sparse image reconstruction algorithms, such as ADMM, is that they can allow direct reconstruction of an accurate sky model, unlike CLEAN based algorithms that produce a restored image \citep{LP18}.

\subsection{$w$-projection $w$-stacking measurement operator}						
In the MPI $w$-stacking $w$-projection algorithm  the measurement operator corrects for the average $w$-value in each $w$-stack, then applies an extra correction to each visibility with the $w$-projection. Each $w$-stack $\bm{y}_k$ has the measurement operator of
\begin{equation}
	\bm{\mathsf{\Phi}}_k = \bm{\mathsf{W}}_k\bm{\mathsf{GC}}_k\bm{\mathsf{F}}\bm{\mathsf{Z}}\bm{\mathsf{\tilde{S}}}_k\,,
\end{equation}
the gridding correction, $\bm{\mathsf{\tilde{S}}}_k$, has been modified to correct for the $w$-stack dependent effects, such as the average $\bar{w}_k$ or the primary beam
\begin{equation}
	\left[\bm{\mathsf{\tilde{S}}}_k\right]_{ii} = \frac{a_k(l_i, m_i){\rm e}^{-2\pi i \bar{w}_k(\sqrt{1 -l^2_i -m^2_i} - 1)}}{g(l^2_i + m^2_i)\sqrt{1 -l^2_i - m^2_i}}\, .
\end{equation}
We choose no primary beam effects within the stack $a_k(l_i, m_i)$. This gridding correction shifts the relative $w$ value in the stack. This can reduce the effective $w$ value in the stack, especially when the stack is close to the mean $\bar{w}_k$, i.e.\ to the value of $w_i - \bar{w}_k$. This reduces the size of the support needed in the $w$-projection gridding kernel for each stack,
\begin{equation}
	\begin{split}
		\left[\bm{\mathsf{GC}}_k\right]_{ij} =
		[GC]\Big(\sqrt{(u_i/\Delta u - q_{u, j})^2 + (v_i/\Delta u - q_{v, j})^2}\\, w_i - \bar{w}_k, \Delta u\Big)\, .
	\end{split}
\end{equation}
$(q_{u, j}, q_{v, j})$ represents the nearest grid points, and we use adaptive quadrature to calculate

\begin{equation}
	\begin{split}
		[GC]\Big(\sqrt{u_{\rm pix}^2 + v_{\rm pix}^2}, w, \Delta u\Big) =
		\frac{2\pi}{\Delta u ^2}\int_{0}^{\alpha/2} g(r)\\
		\times{\rm e}^{-2\pi iw(\sqrt{ 1 - r^2/\Delta u^2} - 1)}
		J_0\left(2\pi r \sqrt{u_{\rm pix}^2 + v_{\rm pix}^2}\right) r{\rm d}r\, ,
	\end{split}
	\label{eq:analytic_convolution_hankel}
\end{equation}
where $g(r)$ is the radial anti-aliasing filter, $\Delta u$ is the resolution of the Fourier grid as determined by the zero padded field of view, and $(u_{\rm pix},v_{\rm pix})$ are the pixel coordinates on the Fourier grid. More details can be found in Paper I.

For each stack \mbox{$\bm{y}_k \in \mathbb{C}^{M_k}$} we have the measurement equation $\bm{y}_k = \bm{\mathsf{\Phi}}_k\bm{x}$. It is clear that each stack has an independent measurement equation. However, the full measurement operator is related to the stacks in the adjoint operators such that
\begin{equation}
	\bm{x}_{\rm dirty} =  \begin{bmatrix}\bm{\mathsf{\Phi}}_1^\dagger,& \dots, &\bm{\mathsf{\Phi}}_{k_{\rm max}}^\dagger \end{bmatrix} \begin{bmatrix} \bm{y}_1 \\ \vdots \\ \bm{y}_{k_{\rm max}}  \end{bmatrix} = \bm{\mathsf{\Phi}}^\dagger \bm{y}\, .
\end{equation}
We use MPI all reduce to sum over the dirty maps generated from each node. The full operator $\bm{\mathsf{\Phi}}$ is normalized using the power method.

\subsection{Clustering $w$-stacks}
It is ideal to minimize the kernel sizes across all stacks, minimizing the memory and computation costs of the kernel. We develop an MPI $k$-means clustering algorithm which greatly improves performance by reducing the values of $|w_i - \bar{w}_k|^2$ across the $w$-stacks. Each MPI node finds the $w$-stack to which a visibility belongs, updating the cluster centers across all MPI nodes with each iteration. This is then followed by an all-to-all MPI operation to distribute the visibilities to their $w$-stacks. There already exist parallel and distributed $k$-means clustering algorithms for big data \citep{kil99,ag13}. The $k$-means $w$-clustering algorithm is presented in Algorithm \ref{algo:kmeans}. This algorithm is necessary to reduce computation and operating memory when applying the $w$-projection kernels by reducing the support size of each kernel. 

\begin{algorithm*}[t]
	\caption{$k$-means $w$-stacking:\newline
		The $k$-means algorithm sorts the visibilities into clusters ($w$-stacks) by minimizing the average $w$ deviation, $(\bar{w} - w)^2$, within each cluster. The algorithm returns two arrays: $\bm{n}$ is the array of indices that labels the $w$-stack for each visibility; $\bm{\bar{w}}$ is the average $w$ value within each $w$-stack. The algorithm requires a starting $w$-stack distribution $\bm{\bar{w}}^{(0)}$, which we choose to be evenly distributed between the minimum and maximum $w$-values. The algorithm should iterate until $\bm{\bar{w}}^{(t)}$ has converged, which we choose to be a relative difference of $10^{-3}$. Note $p$ is the index of visibility, $q$ is the index for $w$-stacks, and $c$ is the place holder for the minimum deviation for the visibility at index $p$. The ${\rm AllSumAll}(x)$ operation is an MPI reduction of a summation followed by broadcasting the result to all compute nodes.} 
	\label{algo:kmeans}							  
	\begin{algorithmic}[1]
		\small
		\Given{$\bm{\bar{w}}^{(0)}, \bm{n}^{(0)}, w_{\rm total}, n_{\rm total}, \bm{w}_{\rm sum}, w_{\rm count}$}
		\RepeatFor{$t=1,\ldots$}
		\State $\bm{w}_{\rm sum} = \bm{0}$
		\State $\bm{w}_{\rm count} = \bm{0}$
		\RepeatFor{$p=1,\ldots$}
		\State $m := 2 (w_{\rm max} - w_{\rm min})$
		\RepeatFor{$q=1,\ldots $}
		\State $c :=(\bm{\bar{w}}^{(t)}_q - \bm{w}_p)^2$
		\State {\bf if} $c < m$ {\bf then}
		\State \quad $m := c$
		\State \quad $\bm{n}^{(t + 1)}_p = q$
		\State {\bf end if}
		\Until $q > n_{\rm total}$
		\State ${\bm{w}_{\rm sum}}_{\bm{n}^{(t + 1)}_p} = {\bm{w}_{\rm sum}}_{\bm{n}^{(t + 1)}_p} + \bm{w}_p$
		\State ${\bm{w}_{\rm count}}_{\bm{n}^{(t + 1)}_p} = {\bm{w}_{\rm count}}_{\bm{n}^{(t + 1)}_p} + 1$
		\Until $p > w_{\rm total}$
		\RepeatFor{$q=1,\ldots $}
		\State $\bm{\bar{w}}^{(t + 1)}_q = 0$
		\State {\bf if} ${\rm AllSumAll}({\bm{w}_{\rm count}}_q) > 0$ {\bf then}
		\State \quad $\bm{\bar{w}}^{(t + 1)}_q = {\rm AllSumAll}({\bm{w}_{\rm sum}}_q)/ {\rm AllSumAll}({\bm{w}_{\rm count}}_q)$
		\State {\bf end if}
		\Until $q > n_{\rm total}$
		\Until {\bf convergence \normalfont}
	\end{algorithmic}
\end{algorithm*}
\subsection{Conjugate symmetry}
Prior to $w$-stacking with the $k$-means algorithm, conjugate symmetry may be used to restrict the $w$-values onto the positive $w$-domain. The origin of the $w$-effect stems from the 3d Fourier transform of a spherical shell and a horizon window, with the $w$ component probing the Fourier coefficient of the signal along the line of sight. The sky, the horizon window, the spherical shell, and the primary beam can all be interpreted as a real valued signal. This provides a conjugate symmetry between $-|w|$ and $+|w|$, i.e.\
\begin{equation}
    y^*(u, v, -|w|) = y(-u, -v, |w|)\, .
\end{equation}
Properties of noise remain unchanged under conjugate symmetry, meaning that measurements can be restricted to positive $w$, i.e.\ $w \in \mathbb{R}_+$. Other modelled instrumental effects may need to be conjugated, which is only important when they are complex valued signals. In particular, polarized signals, e.g.\ Stokes $Q$, $U$, and $V$, are independent real valued signals. Thus, linear polarization has a slightly different relation
\begin{equation}
    y^*_P(u, v, -|w|) = y_Q(-u, -v, |w|) - iy_U(-u, -v, |w|)\, ,
\end{equation}
suggesting the reflection should be done to the Stokes $Q$ and $U$ visibliities before combination into linear polarization, and then combined with $-i$ rather than $+i$. This combination is important for accurate polarimetirc image reconstruction \citep{pra16}.
\subsection{Distributed ADMM}														As in Paper I, we use the alternating direction method of multipliers (ADMM) algorithm implemented in PURIFY \citep{LP18,pra19c} to solve the optimization problem
\begin{equation}
	\label{eq:l1_analysis}
	\min_{{\bm{x}} \in \mathbb{R}^N}\big\| \bm{\mathsf{\Psi}}^\dagger {\bm{x}} \big\|_{\ell_1}\quad {\rm subject}\, {\rm to} \quad \left \|\bm{y} - \bm{\mathsf{\Phi}} {\bm{x}} \right\|_{\ell_2} \leq \epsilon\, ,
\end{equation}
where $\bm{\mathsf{\Psi}}$ is a wavelet transform, the term $\big\| \bm{\mathsf{\Psi}}^\dagger {\bm{x}} \big\|_{\ell_1}$ is a penalty on the number of non-zero wavelet coefficients, while $\left \|\bm{y} - \bm{\mathsf{\Phi}} {\bm{x}} \right\|_{\ell_2} \leq \epsilon$ is the condition that the measurements fit within a Gaussian error bound $\epsilon$. MPI is used to distribute the wavelet transform and enforce fidelity constraints, in conjunction with $w$-stacking.

PURIFY (version 3.0.1, \citealt{purify}) has been updated to implement the $w$-stacking $w$-projection measurement operator with MPI, $k$-means clustering, and conjugate symmetry to efficiently reduce the effective $w$-value within a compute cluster. We find that the use of conjugate symmetry allows the $k$-means algorithm to increase the density of the $w$-stack locations. This in turn reduces the effective $w$ values that are required to be corrected for by the $w$-projection kernels, and greatly decreases the computational burden of the $w$-projection algorithm in the kernel construction.

\section{Application to MWA observation of Fornax A}	
\label{sec:fornaxa}

We use PURIFY (version 3.0.1, \citealt{purify}) to perform wide-field image reconstruction of an observation of Fornax A taken with the MWA. The observation has a pointing centre of 03h 22m 41.7s -37d 12m 30s, and the integration time is 112 seconds. Fornax A was observed using XX and YY polarizations, with the visibilites transformed into Stokes I. The bandwidth was 30.72 MHz with a central frequency of 184.955 MHz and using 768 channels, which is a standard observational mode for the MWA \citep{prabu15,ord15}. The data reduction, including flagging and calibration, is as per \citet{mck15}.

To perform the reconstruction we use 50 nodes of the Grace computing cluster at University College London. Each node of Grace contains two 8 core Intel Xeon E5-2630v3 processors (16 cores total) and 64 Gigabytes of RAM.\footnote{More details can be found at \url{https://wiki.rc.ucl.ac.uk/wiki/RC_Systems\#Grace_technical_specs}}

The reconstructed image is of 2048 by 2048 pixels, with a pixel width of 45 arc-seconds and a field of view of 25 by 25 degrees. The $w$ values range between 0 and approximately 600 wavelengths for the total of 126.6 million visibilites, after conjugating the visibilities for negative $w$ values, i.e.\ a range of 1200 wavelengths originally.

Sorting the visibilities into 50 $w$-stacks (one per MPI node) took under 5 seconds using the MPI distributed $k$-means algorithm described in Algorithm \ref{algo:kmeans}. If the average relative difference of each $w$-stack centre $\bm{\bar{w}}_i$ between $k$-means iterations is less than $10^{-3}$ we consider the algorithm has converged. We do not expect the $w$-projection algorithm performance to improve beyond this level of accuracy in clustering as a function of the number of iterations. In this case, the algorithm converged in 6 iterations.

It took a total of 15 minutes to construct a $w$-projection kernel for all visibilities, using quadrature accuracy of $10^{-6}$ in relative and absolute error, as described in Paper I. The $w$-projection kernel construction time in Paper I was 40 minutes for 50 $w$-stacks (over 25 compute nodes), with the same field of view and same image size, over the same range of $w$ values, but for only 17.5 million visibilities. We find that the use of conjugate symmetry before the $k$-means clustering algorithm allows for more efficient computation of the $w$-projection kernels due to more efficient $w$-stacking because of the reduced range of $w$-values,  allowing for 2.6 times faster kernel construction for approximately 7 times as many measurements (126.6 million visibilities), i.e.\ an overall saving of approximately 18 times.

Reconstruction time took 12 hours, with a total of 2475 iterations, with the FFT and wavelet operations contributing to much of this time due to the large image size. Note that we elected to run the reconstruction for a much longer time than needed to produce an acceptable image. We erred on the side of a higher number of iterations than strictly necessary in order to get a very high quality reconstruction.

\begin{figure*}
	\begin{minipage}{1.0\textwidth}
		\center
		\includegraphics[width=6.5cm]{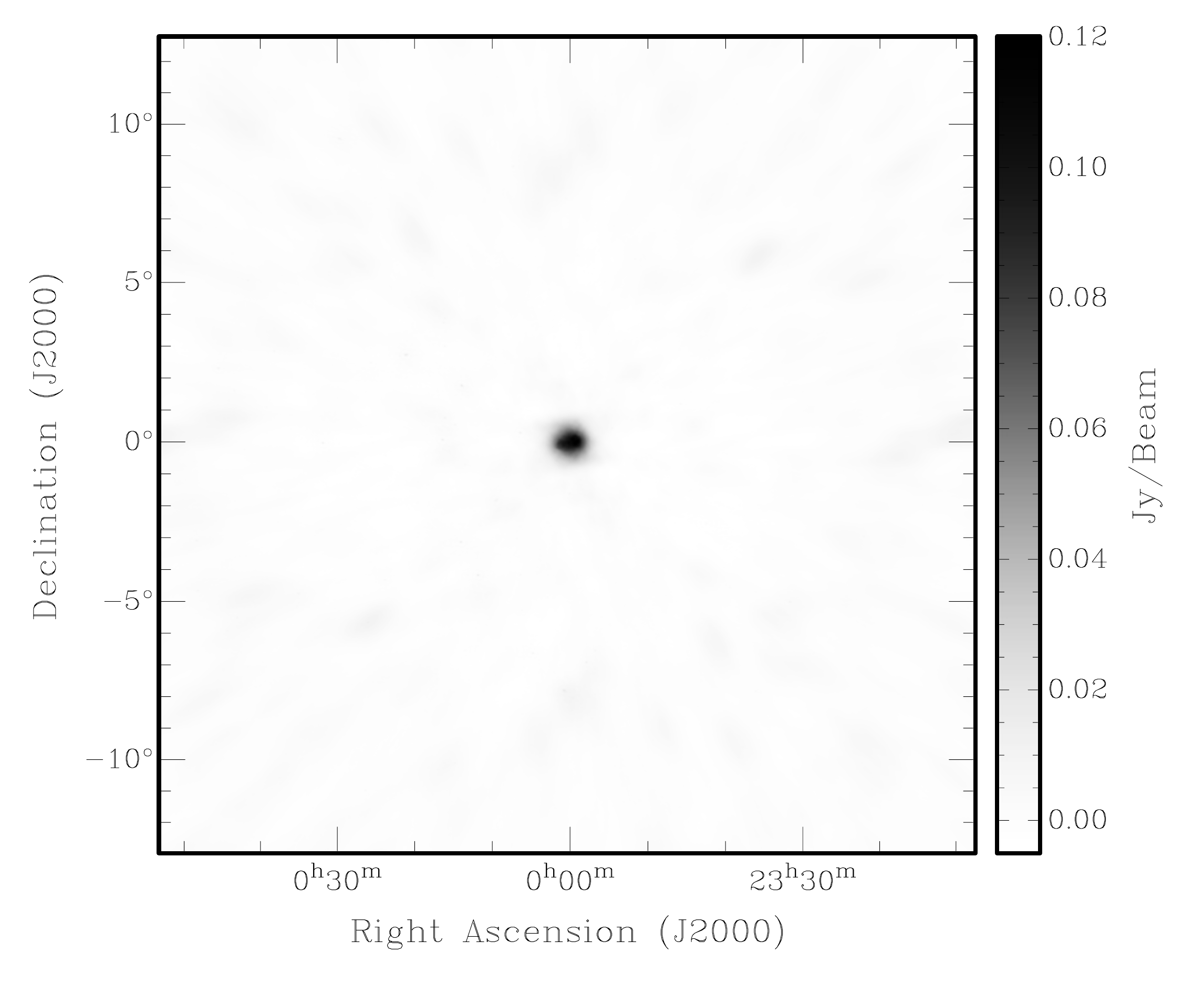}
		\includegraphics[width=6.5cm]{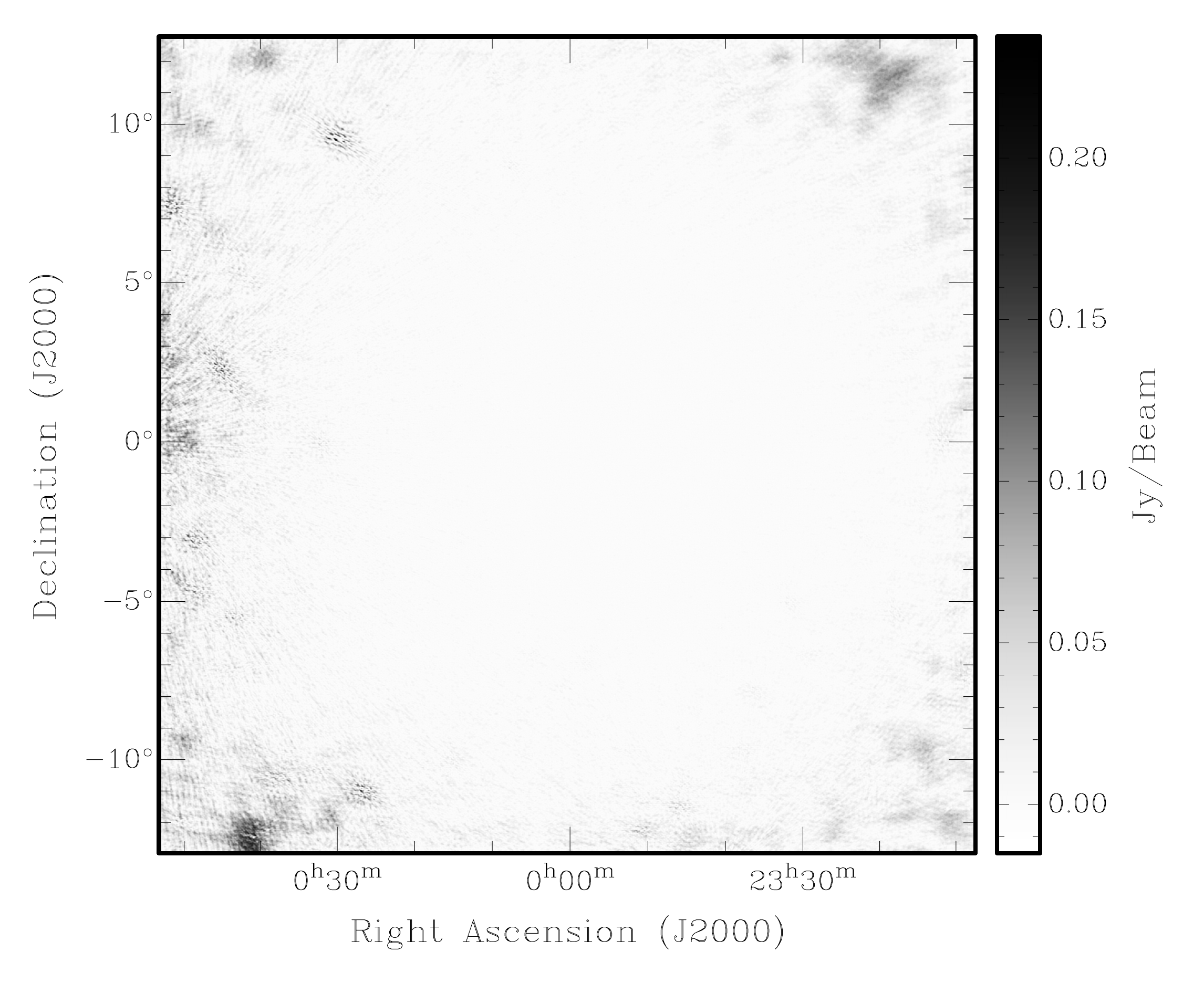}
		\includegraphics[width=16cm]{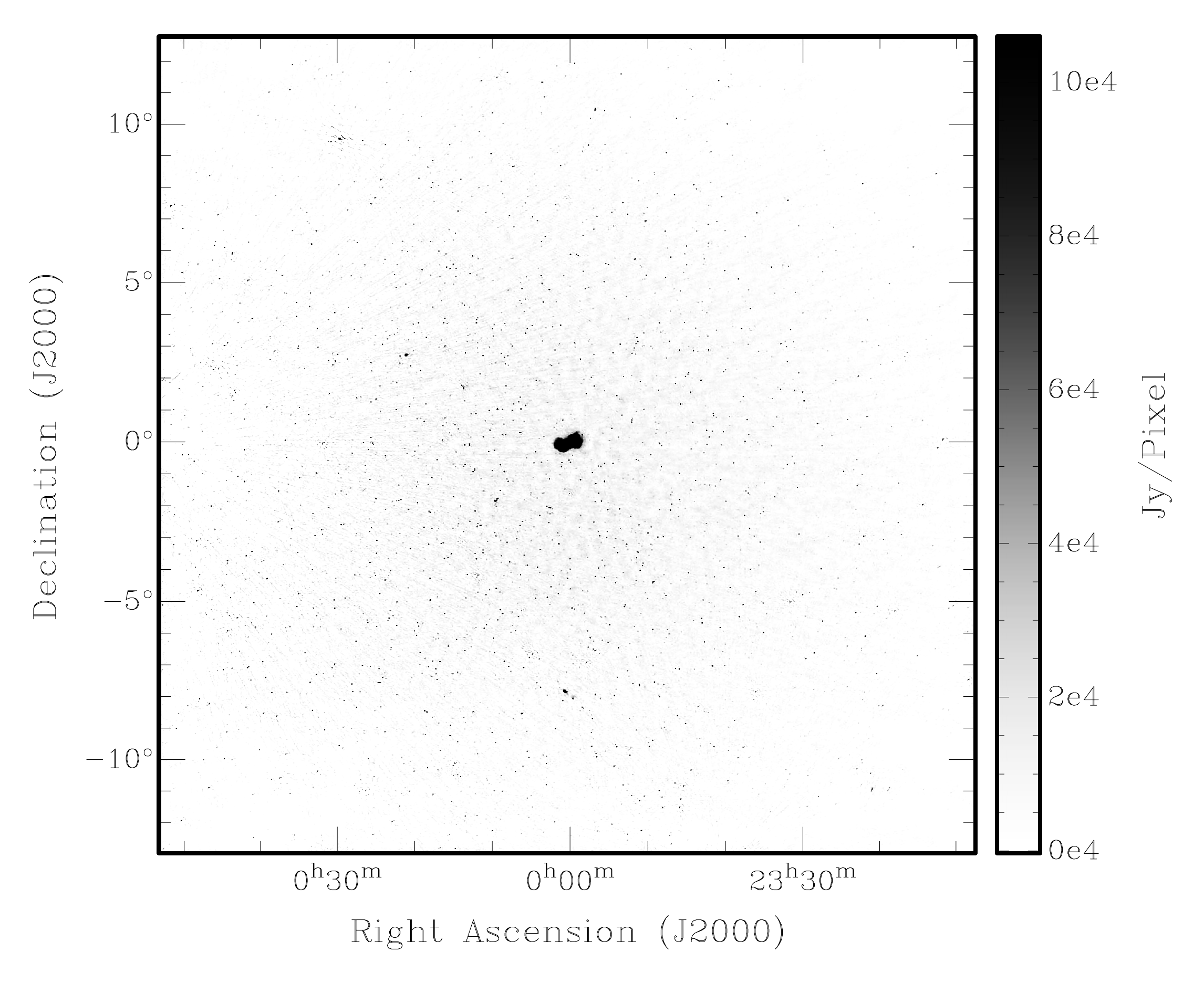}
		\caption{The dirty map (Top Left), residuals (Top Right), and sky model reconstruction (Bottom) of the 112 second MWA Fornax A observation centered at 184.955 MHz, using 126.6 million visibilities and an image size of $2049^2$ (each pixel is 45 arcseconds and the field of view is approximately 25 by 25 degrees). This image was reconstructed using the MPI distributed $w$-stacking-$w$-projection hybrid algorithm, exploiting conjugate symmetry and the $k$-means clustering algorithm for distribution of $w$-stacks presented herein, and using the radial symmetric $w$-projection kernels, in conjunction with the ADMM algorithm. The dynamic range of the reconstruction is 844,000. The RMS of the residuals is approximately 15 mJy/beam over the entire field of view.}
		\label{fig:FornaxA}
	\end{minipage}
\end{figure*}

The reconstructed image can be seen in Figure \ref{fig:FornaxA}, which also shows the residual and dirty maps. The bright, extended source Fornax A is visible at the field centre, with the rest of the field consisting mostly of point sources. The residual map shows that the reconstruction models many of the sources in the field of view, however, the point spread function from bright sources outside the region imaged are still present in the residuals. Despite outside sources disrupting the reconstruction, the root mean squared (RMS) value of the residual map is 15 mJy/beam, and the dynamic range of the reconstruction (as calculated in \citealp{LP18}) is 844,000.

\begin{figure*}
	\begin{minipage}{1.0\textwidth}
		\center
		\includegraphics[width=6.3cm]{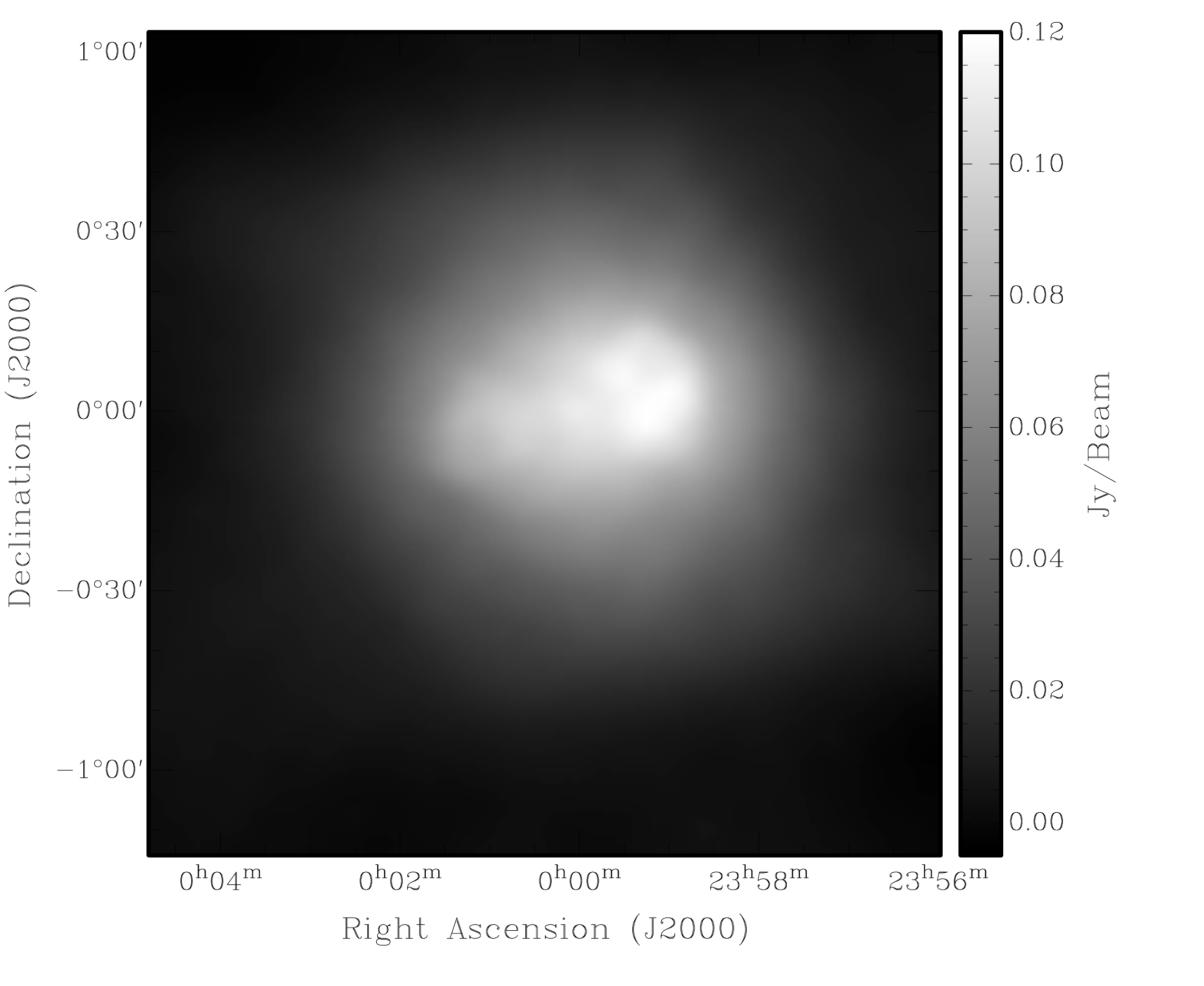}
		\includegraphics[width=6.3cm]{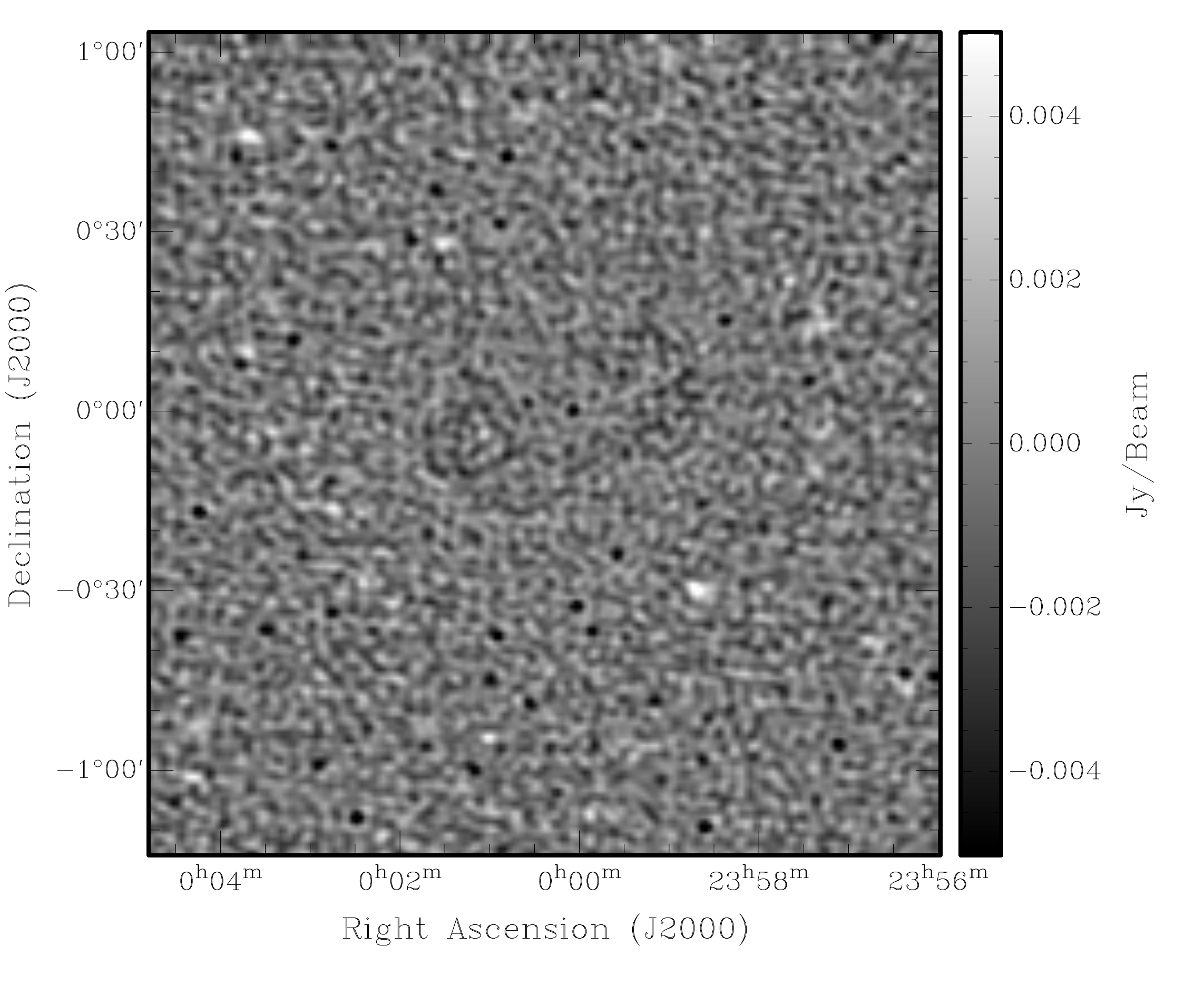}
		\includegraphics[width=16cm]{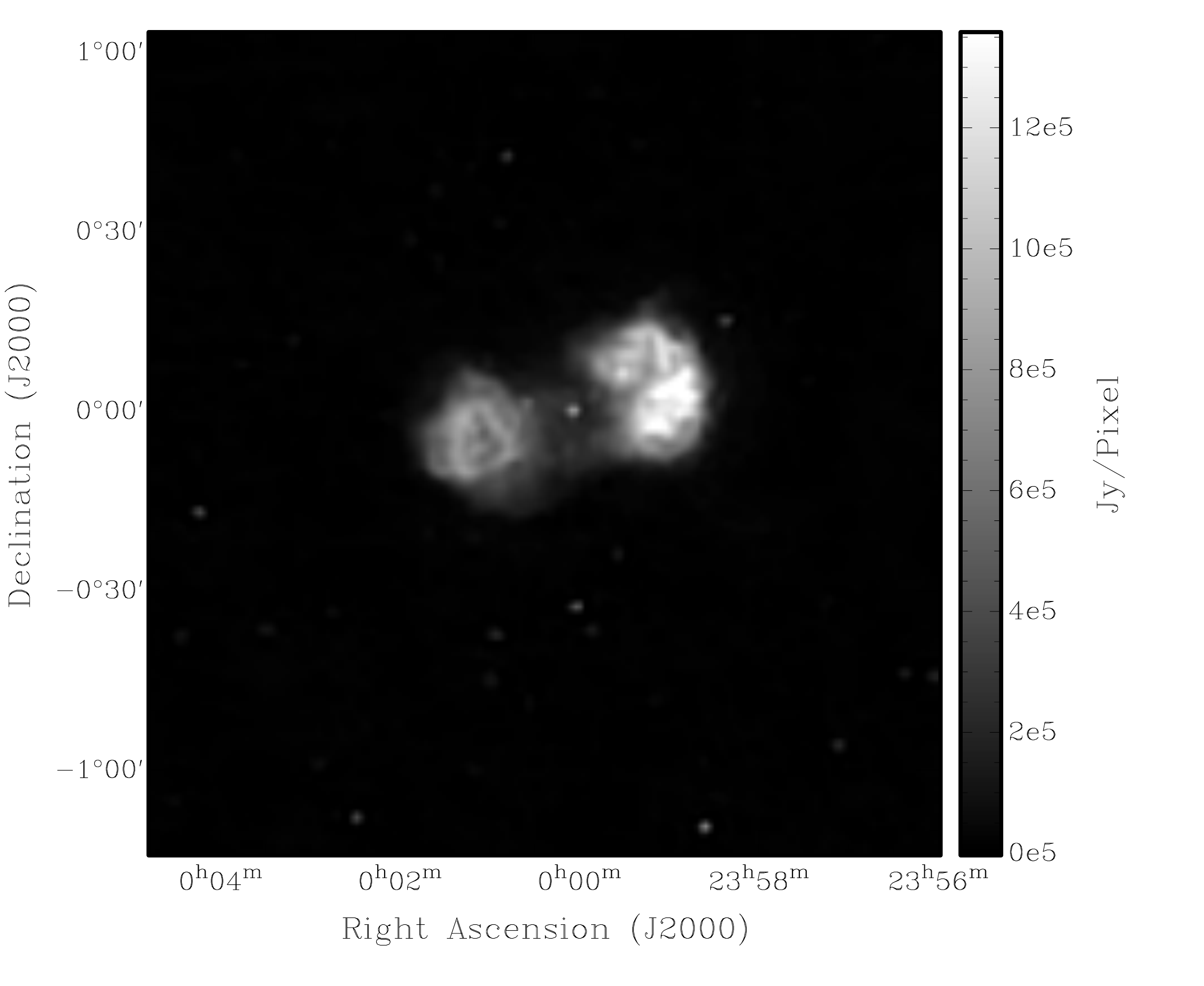}
		\caption{Same as Figure \ref{fig:FornaxA} zoomed view centered on Fornax A, showing the recovered structure of the double lobed radio galaxy. The residuals have been scaled to show the details. The residuals over the zoomed region have an RMS of 1.2 mJy/beam.}
		\label{fig:FornaxA_Zoom}
	\end{minipage}
\end{figure*}

Figure \ref{fig:FornaxA_Zoom} shows a zoom in of Figure \ref{fig:FornaxA}, with the colour scale adjusted to show the reconstruction of Fornax A in greater detail. From the scaled residuals it is clear that this reconstruction accurately models the extended structure of Fornax A.

		
\section{Improvements for the Future}
\label{sec:improvements}
We discuss two classes of possible improvements: kernel interpolations and correction for non-standard direction dependent effects.

\subsection{Kernel interpolation}
\label{sec:kernel_interp}
While we have shown that the use of $k$-means clustering and complex conjugation can aid in kernel construction, $w$-projection kernels can still be expensive in construction time due to the large number of coefficients in $\bm{\mathsf{G}\mathsf{C}}$. This construction overhead can be further reduced using interpolation methods, such as bilinear interpolation between 1d $w$-planes, or parametric fitting. This may allow for on the fly calculation of kernels during imaging. We discuss how a radially symmetric kernel could affect such methods in the future.
																													
\subsubsection{$w$-planes: bilinear interpolation}
The radially symmetric kernel allows fast and accurate calculation, while reducing the dimensions of the kernel. This allows for fast and accurate pre-sampling of the $w$-projection kernel directly in the $uvw$-domain, in some cases to a sufficient pre-sampling density that the error from linear interpolation is negligible compared to the aliasing error. While the mathematical basis for bilinear interpolation is discussed in detail in Paper I, here we present the implementation considerations. 
																													
First we make it clear that a non-radially symmetric kernel would mean pre-sampling in $(u_{\rm pix}, v_{\rm pix}, w)$, which is a computational challenge. For $N_u \times N_v$, samples in $(u, v)$, we would have $N_w$ $w$-projection planes. This requires in total $N_uN_vN_w$ samples. The total memory required in pre-samples is $16\times 10^{-6} \times N_uN_vN_w$[Megabytes]. 
																											
With radial symmetry, we show in Paper I that the $w$-projection kernel can be computed as a function of $(\sqrt{u_{\rm pix}^2 + v_{\rm pix}^2}, w)$. For $N_{uv}$ radial samples in $\sqrt{u_{\rm pix}^2 + v_{\rm pix}^2}$, and $N_w$ samples in $w$, we have only $N_{uv}N_w$ samples. This can be thought of as pre-computing 1d $w$-planes, rather than 2d $w$-planes. Additionally, each sample only requires a 1d integral by quadrature, reducing the pre-sampling time. 
					
The 1d nature of the problem suggests better scaling of pre-sampling computation time and memory, allowing extremely accurate $w$-projection kernels. The total memory required in pre-samples is $16\times 10^{-6} \times N_{uv}N_w$[Megabytes].
					
It is also worth noting that pre-sampling is only required for positive $(u, v, w)$, since the complex conjugate can be used to estimate $(u, v, -w)$ and radial symmetry can be used for negative $u$ and $v$. This leads to additional memory savings in pre-sampling.

Pre-sampling can be optimized for accuracy and storage by using an adaptive sampling density. The pre-samples could be stored permanently in cases where kernel construction is performed repetitively.
					
Bilinear interpolation is computationally cheap, and could make accurate on-the-fly construction of $w$-projection kernels possible, which could be needed for large data such as for the Square Kilometre Array (SKA) \citep{hol17}. In the case where storing the gridding kernels consumes more memory than the pre-sampled kernel, on-the-fly construction can be built into the $\bm{\mathsf{G}\mathsf{C}}$ operator, where bilinear interpolation is used on application. However, memory layout of the pre-samples would be important, since the sample look-up time could reduce the speed of the calculation considerably. 
							
\subsubsection{Function fitting}
Another powerful solution to improve kernel construction costs can be found from the well-known prolate spheroidal wave function (PSWF) gridding kernels, which do not have an analytic form. 
					
PSWFs can be defined multiple ways, such as having optimal localization of energy in both image and harmonic space, making them difficult to compute. They can be calculated directly through Sinc interpolation after solving a discrete eigenvalue problem, but this can be computationally expensive, or they can be calculated using a series expansion. However, this has not stopped radio astronomers using the PSWFs for decades, ever since the work of \cite{sch78a,sch80} described a custom made PSWF that has been used in CASA \citep{mcm07}, AIPS \citep{gre03}, MIRIAD \citep{sault95}, and PURIFY \citep{car14}. In \cite{sch78a,sch80}, a rational approximation is used to provide a stable and accurate fit to the PSWF, which has stood the test of time. 
					
A similar approach can be used to provide an accurate fit to $w$-projection kernels. Put simply, it is possible to fit a radially symmetric kernel as a function of three parameters $\left(\sqrt{u_{\rm pix}^2 + v_{\rm pix}^2}, w, \Delta u\right)$, i.e.\ polynomial fitting. This has various advantages over the pre-sampling method, such as reduced storage, no pre-sampling time, and reduced look up time (which could be critical for on-the-fly application). However, stability and reliability of the fit is not guaranteed and would require further investigation.

\subsection{Additional direction dependent effects}
\label{sec::DDEs}
The 1d radially symmetric kernel framework can be used in conjunction with general 2d kernels that model DDEs.
It is clear that the 1d $w$-projection kernel derivation can be extended to other analytic radially symmetric baseline dependent effects, i.e.\ a function of $r$ or $\sqrt{u^2 +v^2}$ only. But this does not stop the inclusion of more general baseline dependent effects, such as the spectral and polarimetric primary beams and time dependent ionospheric models. Generating these models will require computation that may or may not be worse than the non-coplanar baseline effects, which are telescope dependent. Non-coplanar baseline effects are a special case, where the effects need to be modeled on each baseline and can be modeled in stacks of visibilities. However, in many cases DDE models are station dependent, suggesting the computation is not as extreme as the non-coplanar case. Additionally, these effects may apply to groups of visibilities in time, frequency, and polarization, reducing the number of effects that need to be modeled. 

In the worst case scenario, each baseline will have different DDEs, which can be included by further convolutions (since convolution is commutative)
\begin{equation}
\begin{split}
	[GC](\sqrt{u_{\rm pix}^2 + v_{\rm pix}^2}, w) \to\quad\quad\quad\quad\quad\quad\\ D_{ij}(u, v, w) \star [GC](\sqrt{u_{\rm pix}^2 + v_{\rm pix}^2}, w)\, ,
	\end{split}
\end{equation}				
where $D_{ij}(u, v, w)$ is a model of the DDEs in the $uvw$-domain between two stations $ij$. Typically if $D(u, v, w)$ is band limited, the additional convolution can be performed with a discrete convolution, since $[GC](\sqrt{u_{\rm pix}^2 + v_{\rm pix}^2}, w, \Delta u)$ is also smooth. The discrete convolution has computational complexity $\mathcal{O}(J_{GC}^2J_{D}^2)$, where $J$ is the width of each kernel. If $D$ is separable in $(u, v)$, then this can be reduced greatly to $\mathcal{O}(J_{GC}^2J_{D})$.

The computation of $D(u, v, w)$ may require modeling in the image domain with an FFT for each baseline or it may be known analytically in $(u, v, w)$. In the case where $D_{ij}(u, v, w) = D_j(u, v, w)\star D_i(u, v, w)$ is separable into station dependent effects, it greatly reduces the modeling computation from $N_{\rm Ant}(N_{\rm Ant} - 1)/2 \to N_{\rm Ant}$ kernel constructions. 

The $w$-stacking distribution structure can be applied to model other effects, such as time dependent primary beam and ionospheric models. Distributing the visibilities into (time) $t$, (frequency) $\nu$, and (polarization) $p$ DDE-stacks could alleviate some of the challenges of $D \star GW$ construction; this applies whenever a DDE can naturally be applied to a group of baselines. For a given DDE-stack, we can apply the stack's DDE model directly in the image domain. This can be efficiently done using recent developments in the work of \citet{van18}.

\section{Conclusion}
\label{sec:conclusion}
We have discussed details of the $w$-stacking $w$-projection algorithm implementation, including details of the $k$-means clustering, introduction of conjugate symmetry to improve the computational efficiency of the current algorithm, and possible extensions to the current algorithms and code base to further improve efficiency and accuracy of the reconstructions.  

We use the MPI distributed ADMM
implementation in PURIFY to reconstruct an MWA observation of Fornax A, recovering accurate sky models of the complex source Fornax A and of point sources over the entire 25 by 25 degree field of view. We find that we can construct $w$-projection kernels for 7 times the number of measurements, 2.6 times faster than the time taken in Paper I (an overall saving of approximately 18 times), using the same image size, field of view, and range of $w$ values.

We conclude the work with proposals to modify the implementation of the 1d radial $w$ projection kernels for large data sets, such as the use of kernel interpolation and the inclusion of non radially symmetric directional dependent effects. Accurate correction of wide-field and instrumental effects is critical in the era of next generation radio interferometers and are vital to achieving science goals ranging from the detection of the Epoch of Reionisation to accurately reconstructing cosmic magnetic fields.

% unnumbered section
\section*{Acknowledgements}
We thank Dr Benjamin McKinley for providing the calibrated MWA data of Fornax A. 
This work was supported by the UK Engineering and Physical Sciences Research Council (EPSRC, grants EP/M011089/1).  The authors acknowledge the use of the UCL Grace High Performance Computing Facility (Grace@UCL), and associated support services, in the completion of this work. 

{\it Facilities:} MWA																																
\bibliographystyle{pasa-mnras}
\bibliography{refs}

\end{document}